\documentclass[aps,preprint,showpacs,showkeys,titlepage,amsmath,amssymb,amsfonts]{revtex4-2}
\usepackage{xcolor}
\usepackage{graphicx} 

\begin{document}


\title{A chiral model for sterile neutrino}

\author{Chun Liu$^{1, 2}$} \email[Email: ]{liuc@mail.itp.ac.cn} \quad
\author{Yakefu Reyimuaji$^{1, 3}$}\email[Email: ]{yreyi@hotmail.com (corresponding author)}

\affiliation{$^1$ CAS Key Lab. of Theor. Phys., 
Institute of Theoretical Physics, 
Chinese Academy of Sciences, Beijing 100190, China\\
$^2$ School of Physical Sciences, UCAS, 
Chinese Academy of Sciences, Beijing 100049, China\\ 
$^3$ School of Physical Science and Technology, Xinjiang University, Urumqi, Xinjiang 830046, China}


\date{\today}

\begin{abstract}
A model, which extends the standard model with a new chiral U(1)$'$ gauge symmetry sector, for the eV-mass sterile neutrino is constructed. It is basically fixed by anomaly free conditions. The lightness of the sterile neutrino has a natural explanation. As a by product, this model provides a WIMP-like dark matter candidate. 
\end{abstract}



\maketitle


\section{I\lowercase{ntroduction}} 
\label{sec:intro}

There are several compelling anomalies in different neutrino experiments, which cannot be explained by the standard three-neutrino oscillation. LSND experiment has reported an excess in the $\bar{\nu}_e$ appearance from $\bar{\nu}_\mu$ beam with $3.8\sigma$ significance \cite{LSND:2001aii}. MiniBooNE has confirmed the result and reported the excess with $2.8\sigma$ significance in $\bar{\nu}_\mu \to \bar{\nu}_e$ channel and $4.5\sigma$ significance in $ \nu_\mu \to \nu_e$ channel, combination of these two channels reaches $4.7\sigma$ \cite{MiniBooNE:2018esg}. It is worth noticing that the combined significance of excesses from the LSND and MiniBooNE experiments reaches to as large as 6$\sigma$ \cite{MiniBooNE:2018esg}. Another anomalous result has arisen in reactor experiments, the observed electron antineutrino rate showed a deviation about $6\%$  from theoretical expectation with a significance of $2.8\sigma$ \cite{Mueller:2011nm, Huber:2011wv}. In addition, about a $15 \%$ rate deficit of electron neutrinos is observed with a $3.0\sigma$ significance in calibration of radiochemical experiments \cite{GALLEX:1997lja, SAGE:1999nng, SAGE:2009eeu, Kaether:2010ag}, which use inverse beta decays on the Gallium. 

Even though these anomalies are rather old, reasons behind them are still unclear, and we do not know if they stem from a common origin. Yet there are intriguing works on the possibility of explaining these anomalies by introducing a sterile neutrino on top of three active neutrinos, and on determination of the parameter space allowed by experiments (for reviews, see~\cite{Gariazzo:2017fdh, Boser:2019rta, Diaz:2019fwt, Bilenky:2019gzn}). Results from such a paradigm indicate that the sterile neutrino has a mass about $1$ eV and mixes with the electron neutrino at about 1$\%$ level. If this is true, it will have  profound  implications in searching for new physics beyond the Standard Model (SM). At this point, we also want to mention that Daya bay and MINOS+ collaborations announced that their results exclude most of the sterile neutrino parameter space allowed by LSND and MiniBooNE experiments at $99\%$  confidence level \cite{MINOS:2020iqj}. Meanwhile, RENO collaboration performed model independent search for sterile neutrino oscillation by using RENO and NEOS data, and claimed that they obtained $68\%$ CL allowed region with the best fit values of $\left| \Delta m^2_{41} \right| = 2.37 \ \pm 0.03 \ \text{eV}^2$ and  $\sin^2 2\theta_{14} =0.09\pm0.03$~\cite{Atif:2020glb}. These claims may be further solidified by upcoming data from the experiments and, at the same time, tested by other experiments. Although there are many experimental efforts, no conclusive evidences come out to confirm or rule out the interpretation of these anomalies in terms of the oscillations between the active and sterile neutrino states.  

Studies about the framework of three active and one eV-mass sterile neutrino have drawn considerable attention. For instance, impacts of the sterile neutrino on the active neutrino masses and mixing, on the long- and short-baseline experiments including matter effects, and on the CP asymmetry have been carried out \cite{Reyimuaji:2019wbn, Branco:2020yvs}. Furthermore, how to extend the SM to that including one sterile neutrino with an eV-scale mass is an interesting question. In the present paper, we try to understand the origin of the eV-scale sterile neutrino mass from the point of view of simplicity, by adding a new U(1) gauge sector. Its lightness is attributed to that the sterile neutrino is chiral, meaning that none of the charges of newly introduced  fermions under the U(1) is zero, nor opposite to each other. These U(1) charges are subject to anomaly cancellation conditions.  As we will show, a small sterile neutrino mass can be successfully generated and the mixing can be achieved.

\section{T\lowercase{he model}} 
\label{sec:model}

We introduce a new U(1) gauge interaction in which the sterile neutrino  is a chiral fermion.  To distinguish it from the SM $U(1)_\mathrm{Y}$, this symmetry is denoted as U(1)$'$.  The model is largely fixed by anomaly free conditions \cite{Costa:2019zzy,Geng:1989tcu,He:1990me,Kong:1996as,deGouvea:2015pea,Wong:2020obo}.  Because the SM has been very successful experimentally, we assume the fields of the SM are neutral under the U(1)$'$.  In addition, the right-handed neutrinos 
which are for the tiny masses of SM neutrinos are also assumed neutral under the U(1)$'$.  The simplest chiral U(1)$'$ model is generally studied in Ref. \cite{Costa:2019zzy}.  The smallest number of fermionic fields which are nontrivial under the U(1)$'$ is five.  Our model assumes that there are 5 chiral fermion fields $\chi_1$, $\chi_2$, ..., $\chi_5$ which carry nonvanishing U(1)$'$ charges $z_1$, $z_2$, ..., $z_5$, respectively. They satisfy 
\begin{equation}
\sum_{i=1}^5 z_i^3 = 0 \,,~~~\sum_{i=1}^5 z_i = 0 \,.  
\end{equation}  
We further introduce three Brout-Englert-Higgs fields of U(1)$'$, $\phi_1$, $\phi_2$ and $\phi_3$ with U(1)$'$ charges $\tilde{z}_1 = -(z_1+z_3)$, 
$\tilde{z}_2 = -(z_2+z_4)$ and $\tilde{z}_3 = -z_5$, respectively.  To be specific, the charges are chosen as the following, 
\begin{equation}
z_1=\frac{1}{9}\,,~z_2=\frac{5}{9}\,,~z_3=-\frac{7}{9}\,,~z_4=-\frac{8}{9}\,,~z_5=1\,,~
\end{equation} 
thus $\phi_i (\tilde{z}_i)$'s are $\phi_1(\displaystyle\frac{2}{3})$, $\phi_2(\displaystyle\frac{1}{3})$, and $\phi_3(-1)$.  In terms of two-component Weyl spinors, the relevant renormalizable Lagrangian is 
written as the following, 
\begin{equation}
\label{lagrangian}
\begin{array}{lll}
\mathcal{L}_{\rm U(1)'} &=& -\frac{1}{4}F'_{\mu\nu}F'^{\mu\nu}
-i\bar{\chi}_i\bar{\sigma}^\mu D_\mu \chi_i-(D_\mu\phi_i)^*(D^\mu\phi_i) \\ 
&&-(c_1\chi_1\chi_3\phi_1+c_2\chi_2\chi_4\phi_2+c_3\chi_1\chi_2\phi_1^*+{\rm h.c.})\\
&&-V(\phi_1, \phi_2, \phi_3) \,, 
\end{array} 
\end{equation} 
where the U(1)$'$ covariant derivative $D_\mu =\partial_\mu-i\tilde{z}_i g'A'_\mu$ with $A'_\mu$ being the U(1)$'$ gauge field, and  $\tilde{z}_i$ is U(1)$'$ charge of $\phi_i$. Here $g'$ and $c_i$'s stand for coupling constants, $V(\phi_1, \phi_2, \phi_3)$ is the potential of scalar fields. Its expression is
\begin{equation}\label{newspot}
\begin{aligned}
 V(\phi_1, \phi_2, \phi_3, h)  = & \sum_i \left( \frac{\kappa_i}{2} \left| \phi_i \right|^4+ \alpha_i \left| h \right|^2 \left| \phi_i \right|^2 -\omega^2_i \left| \phi_i \right|^2 \right) \\
 & + \left( \beta_1\phi_1 \phi_2 \phi_3 + \beta_2\phi^*_1 \phi^2_2  + \beta_3 \phi^3_2 \phi_3 + \beta_4 \phi^2_1 \phi^*_2 \phi_3+{\rm h.c.}\right),
 \end{aligned}
\end{equation}
where interaction with the SM higgs $h$ is taken into consideration. The full potential of the model is a sum of that in eq.~\eqref{newspot} and the higgs potential in the SM,
\begin{equation} \label{eq:higgspotent}
	V(h) = -\mu^2 h^\dagger h +\lambda (h^\dagger h)^2.
\end{equation}
 The potential in eq.~\eqref{newspot} is rather simple, which includes 5 dimensionful and 8 dimensionless parameters. At the moment we do not specify these parameters, but just assume that spontaneous breaking of the U(1)$'$ gauge symmetry can be triggered. 

When scalar fields $\phi_i$ get vacuum expectation values (VEVs) $\langle\phi_i\rangle$, first four fermions $\chi_1$-$\chi_4$ become massive, their mass matrix arises from eq.~\eqref{lagrangian},
\begin{equation}\label{mamatchi}
 \mathcal{M}_\chi = \frac{1}{2} \begin{pmatrix}
                     0 & m^\chi_3 & m^\chi_1 & 0\\
                      m^\chi_3  & 0 & 0 & m^\chi_2 \\
                      m^\chi_1 & 0 & 0 & 0   \\
                     0 & m^\chi_2 & 0 & 0
                    \end{pmatrix},
\end{equation}
where $m^\chi_1 = c_1\langle\phi_1\rangle $, $m^\chi_2 = c_2 \langle\phi_2\rangle$, and $m^\chi_3 = c_3 \langle\phi_1\rangle^*$. Diagonalization of this mass matrix shows that there are two pairs of degenerate masses
\begin{equation}
 \begin{aligned}
  m^\chi_\pm = \frac{1}{2\sqrt{2}} \left[ \sum^3_{i=1} \left| m^\chi_i\right|^2 \pm \left( \left( \sum^3_{i=1} \left| m^\chi_i\right|^2\right)^2 -4\left| m^\chi_1\right|^2\left| m^\chi_2\right|^2\right)^{1/2}\ \right]^{1/2}.\\
 \end{aligned}
\end{equation}
Above result implies that the smaller mass $ m^\chi_-$ would vanish if either $ m^\chi_1$ or $ m^\chi_2$ were zero. This is expected from eq.~\eqref{mamatchi} since $\det \left( \mathcal{M}_\chi \right) = \left(m^\chi_1 m^\chi_2/4  \right)^2$. The mass-squared difference, $\left(m^\chi_+ \right)^2 - \left(m^\chi_- \right)^2$, has a minimum value $\frac{1}{4} \sqrt{\left| m^\chi_3\right|^4+4\left| m^\chi_1\right|^2\left| m^\chi_3\right|^2 }$ if $\left| m^\chi_1\right| = \left| m^\chi_2\right|$. From this point of view, $ m^\chi_1$ and $ m^\chi_2$ play rather important roles: they not only determine the lower bound of the smaller mass but also that of the mass-squared difference.

The mass eigenstates can be simply written in terms of Dirac fields. These four massive chiral fermions can be taken as two massive Dirac particles, 
\begin{equation} 
\Psi_+ \equiv \left(\begin{array}{c}
\chi'_1\\
\bar{\chi}'_3
\end{array}
\right)\;, ~~~ 
\Psi_- \equiv \left(\begin{array}{c}
\chi'_2\\
\bar{\chi}'_4
\end{array}
\right) \; ,  
\end{equation}  
where $\chi'_1$ and $\chi'_3$ are one of the degenerate pair after diagonalization with mass $m^\chi_+$, and $\chi'_2$ and $\chi'_4$ are another pair with mass $m^\chi_-$. Therefore, the Lagrangian involving $\Psi_+$ and $\Psi_-$ becomes  
\begin{equation}\label{eq:inteact}
\begin{aligned}
\mathcal{L}_{U(1)'} \supset & \ \bar{\Psi}_+ \left(\gamma^\mu i D_\mu -  m^\chi_+ \right)\Psi_+ + \bar{\Psi}_- \left(\gamma^\mu i D_\mu - m^\chi_ - \right)\Psi_- + \frac{1}{2} m^2_A A'^\mu A'_\mu  \\
& + c'_1 \bar{\Psi}_+ P_L \Psi_+ \phi_1 +  c'_2 \bar{\Psi}_- P_L \Psi_- \phi_2 + c'_3 \bar{\Psi}_+ P_L \Psi_- \phi^*_1+{\rm h.c.} \,,
\end{aligned}
\end{equation}
where $m_A$ is the U(1)$'$ gauge boson mass, $ \displaystyle m^2_A=2g'^2 \sum_i \tilde{z}^2_i \left| \langle \phi_i \rangle \right|^2$, $P_L=\left( 1-\gamma_5 \right)/2$ and $c'_i$ ($i=1-3$) stands for the Yukawa couplings in the mass eigenstate basis of TeV-scale massive fermions.

The chiral fermion $\chi_5$ is still massless at the renormalizable interaction level, because it has no Yukawa interactions with any scalar field.  It is this $\chi_5$ which we take as the sterile neutrino.  Its small mass is due 
to dimension-5 operators in the Lagrangian, 
\begin{equation}
\label{dim5lagrangian}
\begin{array}{lll}
\mathcal{L}_{\rm U(1)'}^{\rm dim.=5}&=&-\displaystyle\frac{\lambda_1}{M}\chi_5\chi_5\phi_3^2
-\displaystyle\frac{\lambda_2}{M}\chi_3\chi_4\phi_1\phi_3^*
-\displaystyle\frac{\lambda_3}{M}\chi_1\chi_2 \phi_2\phi_3
-\displaystyle\frac{\lambda_4}{M}\chi_1\chi_3 \phi^*_2\phi^*_3\\
&&-\displaystyle\frac{\lambda_5}{M}\chi_2\chi_4 \phi_1\phi^*_2
-\displaystyle\frac{\lambda_6}{M}\chi_2\chi_4 \phi^*_1\phi^*_3+{\rm h.c.} \,,
\end{array} 
\end{equation} 
where $\lambda_i$'s are couplings, and $M$ is the new physics scale.  After U(1)$'$ symmetry breaking, above terms in the Lagrangian provide  $1/M$ suppressed contributions to $\chi_i$ masses.  These are just small corrections to $\chi_{1-4}$ masses.  It is essential to note that the first term results in a nonvanishing (but small) mass of $\chi_5$, namely it gives $\chi_5$ a Majorana mass, $\displaystyle\sim\frac{\lambda_1\langle\phi_3\rangle^2}{M}$. This is an essence of the model to explain the very small sterile neutrino mass and, at the same time, to generate several heavy particles leaving very small observable effects in current experiments, some of which may play a role of the dark matter candidate, as a bonus.

Considering connection to the SM sector, the U(1)$'$ sector mixes with the SM in the following way, 
\begin{equation}\label{eq;mixingu1&u1p}
\mathcal{L}_{\rm mixing}=\epsilon F^{\mu\nu}F'_{\mu\nu}
+\displaystyle \frac{\lambda_3^i}{M} (h^T i\sigma_2 l_i) \chi_5\phi_3
+{\rm h.c.} \,, 
\end{equation} 
where $F_{\mu\nu}$ with (without) prime is the U(1)$'$ (the SM ${\rm U(1)_Y}$) gauge field strength, $l_i$ stands for the SM lepton doublets ($i=1, 2, 3$). The parameter $\epsilon$ measures the size of kinetic mixing of ${\rm U(1)_Y}$ and the U(1)$'$ gauge sectors. There is a mixing between the sterile and active neutrinos from the second term, and $\lambda_3^i$'s are parameters to quantify the mixing.  After spontaneous breaking of SM and U(1)$'$ symmetries, neutrino mass mixing terms $l_i\chi_5$ appear in the Lagrangian, whose magnitude is $\displaystyle \frac{\lambda_3^i\langle h\rangle\langle\phi_3\rangle}{M}$.   

Numerically, to make the active-sterile neutrino mixing to be about 
$0.1$, it generically requires that the mass ratio 
$\displaystyle\frac{\lambda_3^i\langle h\rangle\langle\phi_3\rangle}{M}/\frac{\lambda_1\langle\phi_3\rangle^2}{M}\sim 0.1$, 
this gives that $\langle\phi_3\rangle\sim$ $100$ GeV  -- $1$ TeV, by taking $\lambda_3^i$ and $\lambda_1$ to be the same order.  Furthermore, taking $\chi_5$ mass to be 1 eV, it follows that $M\sim (10^{12}$ -- $10^{14})$ GeV, namely 
the seesaw scale provided that coupling $\lambda$'s are of natural values $\sim 0.1$.   

For a UV completion of the theory, one can introduce right-handed neutrinos $N_R$'s with masses about $M$, which are supposed to be singlets of both the SM and U(1)$'$ gauge symmetries. This allows to write $\chi_5 N_R \phi_3$ term in the Lagrangian and it, in turn, generates the first term of eq.~\eqref{dim5lagrangian} and the second term of eq.~\eqref{eq;mixingu1&u1p} at tree level after integrating out the heavy right-handed neutrino. 

Let us look at the full $4\times 4$ neutrino mass matrix  
\begin{equation}
\label{15}
{\mathcal M}^\nu = \left(
\begin{array}{cc}
M^\nu_{ij} & m_{is}^\nu \\
m_{is}^{\nu T}         & m^\nu_{ss} \\[3mm] 
\end{array}
\right) \,, 
\end{equation} 
where $M^\nu_{ij}$ stands for the $3\times 3$ active neutrino mass 
matrix, it reads 
\begin{equation}
\label{Wbrg}
M^\nu_{ij} = c_{ij} \frac{\langle h\rangle^2}{M}\, ,
\end{equation} 
with coefficients $c_{ij}$.  It is interesting to notice that we take $M$ as the seesaw scale, and thus
\begin{equation}
\label{twomasses}
\begin{array}{lll}
m^\nu_{ss} &=& \displaystyle\frac{\lambda_1\langle\phi_3\rangle^2}{M} \,, \\
m_{is}^\nu &=& \displaystyle\frac{\lambda^i_3 \langle h\rangle \langle\phi_3\rangle}{M} \,.  
\end{array}
\end{equation} 
The fourth mass eigenstate $\nu_s$ of the matrix \eqref{15}  has a mass $m_s \simeq 1$ eV with a mixing $\left|U_{i4}\right| \simeq 0.1$ with active neutrinos $\nu_i$.

\section{R\lowercase{elevant phenomenology}} 
\label{sec:Pheno}

Let us look at the boson sector. After electroweak and U(1)$'$ symmetries  spontaneous breaking, from eq.~\eqref{newspot}, there appears mass-squared mixing terms between the SM higgs and scalars. To show it quantitatively for illustrative purpose, we consider a possibility of the VEVs aligned along the fields $h$ and $\phi_3$, and expand these fields around the vacuum,
\begin{equation}\label{eq:scalarvevs}
          h = \frac{1}{\sqrt{2}} \begin{pmatrix} 
		0\\ 
		v + H
	\end{pmatrix}, \quad \phi_3 =  \frac{1}{\sqrt{2}} (u_3 + \varphi_3)  e^{i \theta_3}.
\end{equation}
The VEVs are determined by minimization of the scalar potential,
\begin{equation}
	\begin{aligned}
		&v^2 =\frac{2 \left(\kappa_3 \mu ^2 -\alpha_3 \omega^2_3 \right)}{2 \lambda  \kappa_3 -\alpha^2_3}, \\
		& u^2_3= \frac{2 \left(2 \lambda \omega^2_3- \alpha_3 \mu^2 \right)}{ 2 \lambda  \kappa_3 -\alpha^2_3}.
	\end{aligned}
\end{equation}
The mass-squared matrix of  $h$ and $\phi_3$ is 
\begin{equation}
\frac{1}{ 2\lambda \kappa_3 - \alpha_3^2 }
	\left(
	\begin{array}{cc}
	2 \lambda  \left( \kappa_3 \mu^2 - \alpha_3  \omega_3^2\right) & \alpha_3\sqrt{\left(\kappa_3 \mu ^2 - \alpha_3 \omega^2_3 \right) \left(  2 \lambda \omega^2_3  - \alpha_3 \mu^2\right) }\\
    \alpha_3\sqrt{\left(\kappa_3 \mu ^2 - \alpha_3 \omega^2_3 \right) \left(  2 \lambda \omega^2_3  - \alpha_3 \mu^2\right) } & \kappa_3 \left(2 \lambda  \omega_3^2 - \alpha_3 \mu^2 \right) \\
	\end{array}
	\right).
\end{equation}
 Diagonalization of this matrix can be done via $2\times 2$ rotation. The corresponding rotation angle $\delta$ is
\begin{equation}
	\begin{aligned}
		\tan 2\delta =  \frac{2 \alpha_3\sqrt{\left(\kappa_3 \mu ^2 - \alpha_3 \omega_3^2\right) \left(2 \lambda  \omega_3^2 - \alpha_3 \mu ^2\right)}}{\kappa_3 \mu ^2 (\alpha_3 +2 \lambda )-2 \lambda  \omega_3^2 (\alpha_3+\kappa_3)},
	\end{aligned}
\end{equation}
 and mass-squared eigenvalues are as follows
\begin{equation}\label{eq:masssrrd}
\begin{aligned}
	m^2_1 = & \ \frac{\kappa_3 \mu ^2 (2 \lambda - \alpha_3 )+2 \lambda  \omega_3^2 (\kappa_3 - \alpha_3 )+\Omega}{2 \left(2 \kappa_3 \lambda - \alpha_3^2   \right)}, \\
	m^2_2 = & \ \frac{\kappa_3 \mu ^2 (2 \lambda - \alpha_3 )+2 \lambda  \omega_3^2 (\kappa_3 - \alpha_3 )-\Omega}{2 \left(2 \kappa_3 \lambda - \alpha_3^2   \right)},
\end{aligned}	
\end{equation}	
where
\begin{equation}
	\begin{aligned}
	\Omega^2 = & 4 \alpha_3^2 \left(\kappa_3 \mu ^2 - \alpha_3 \omega_3^2\right) \left(2 \lambda  \omega_3^2 - \alpha_3 \mu ^2\right)+\left[\kappa_3 \mu ^2 (\alpha_3 +2 \lambda )-2 \lambda  \omega_3^2 (\alpha_3 +\kappa_3)\right]^2.
    \end{aligned}
\end{equation}	
The mass eigenstate $H'$ with mass $m_1$, which is SM higgs-like, and the mass eigenstate $\varphi'_3$ with mass $m_2$, that is $\varphi_3$-like, are
\begin{equation}
\begin{aligned}
	H' = & \cos \delta \, H - \sin \delta \, \varphi_3, \\
	\varphi'_3 =  &  \sin \delta \, H + \cos \delta \, \varphi_3.
\end{aligned}
\end{equation}

A phenomenological consequence is on the higgs property. The higgs portal interactions modify the higgs boson self-couplings. One can derive a modified trilinear $H'$ coupling,
\begin{equation}
	\frac{1}{2} \left(- \alpha_3 u_3 \sin \delta \cos^2 \delta -\kappa_3 u_3 \sin^3 \delta +\alpha_3 v \sin^2\delta \cos \delta +2 \lambda  v \cos^3\delta \right) H^{'3}.
\end{equation}
From this result, one can easily show that when the mixing is zero it reproduces $\lambda v H^3$,  which is the trilinear SM higgs coupling. Taking $u_3 \sim 1$ TeV, then $v/u_3 \sim 0.1$, the trilinear coupling is about $\left(\lambda -\frac{\alpha^2_3}{\kappa_3} \right) v$. This modification is quite significant, it can be $\sim 100\%$ if all the couplings are of the same order. Precision measurements of higgs couplings are exactly in the goal of higgs factory experiments, such as the Circular Electron Positron Collider (CEPC)~\cite{CEPCStudyGroup:2018ghi}. The trilinear higgs coupling of our model will be very well tested by the CEPC.

\begin{figure}
  \includegraphics[width=0.45\textwidth]{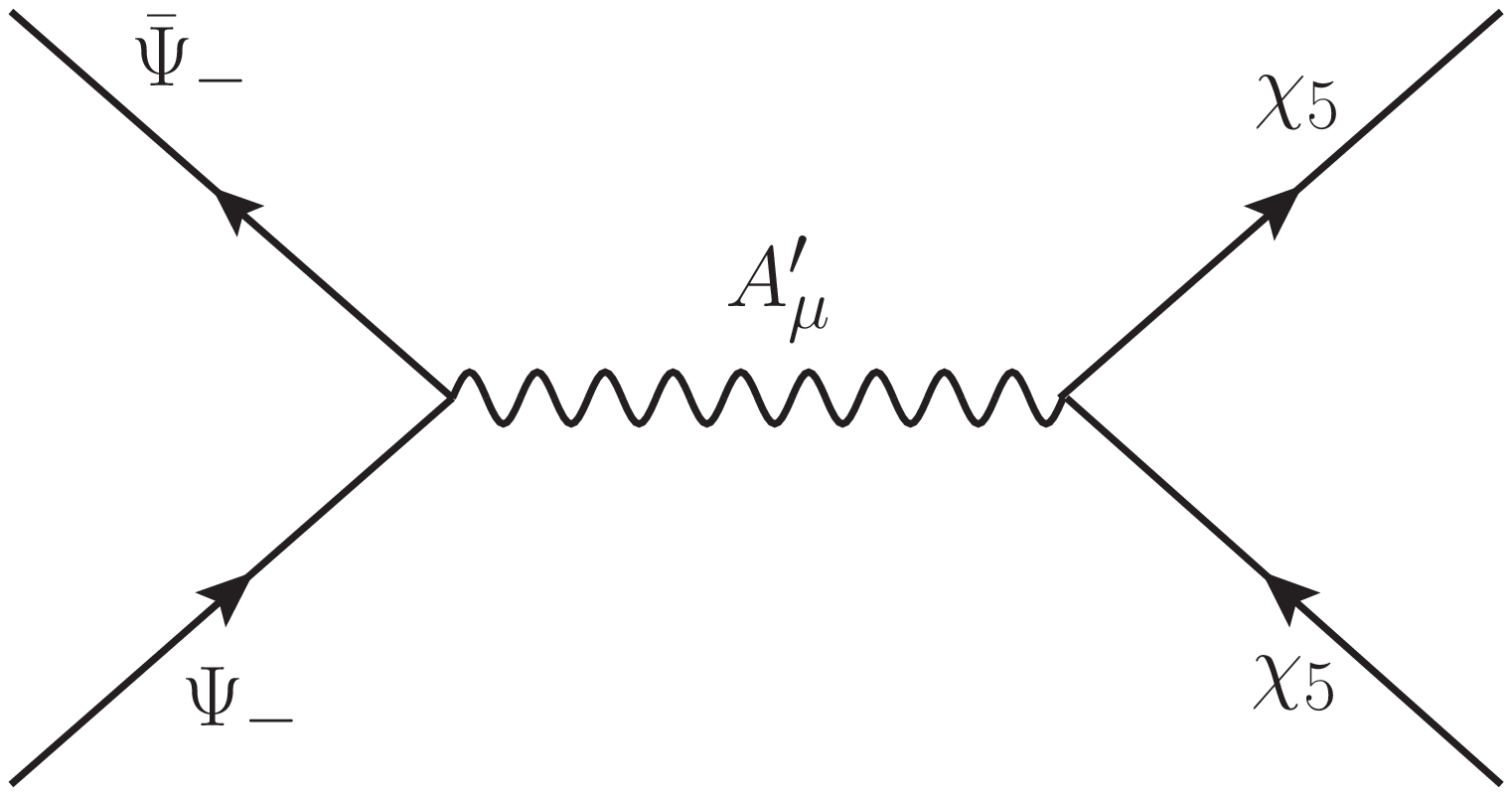} \hspace*{2mm}
  \includegraphics[width=0.45\textwidth]{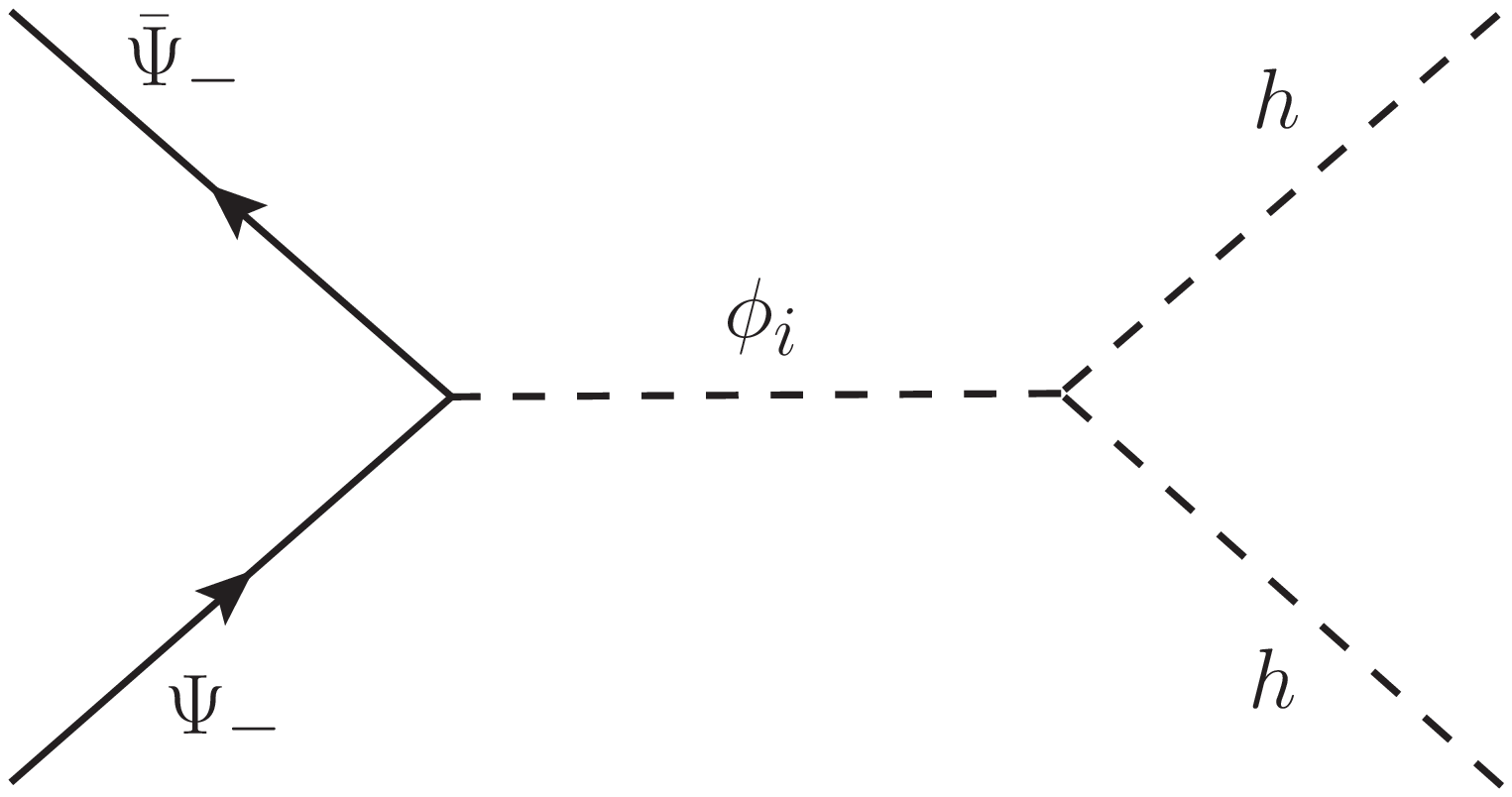}\\
  \vspace*{5mm}
  \hspace*{-0.47\textwidth}
  \includegraphics[width=0.45\textwidth]{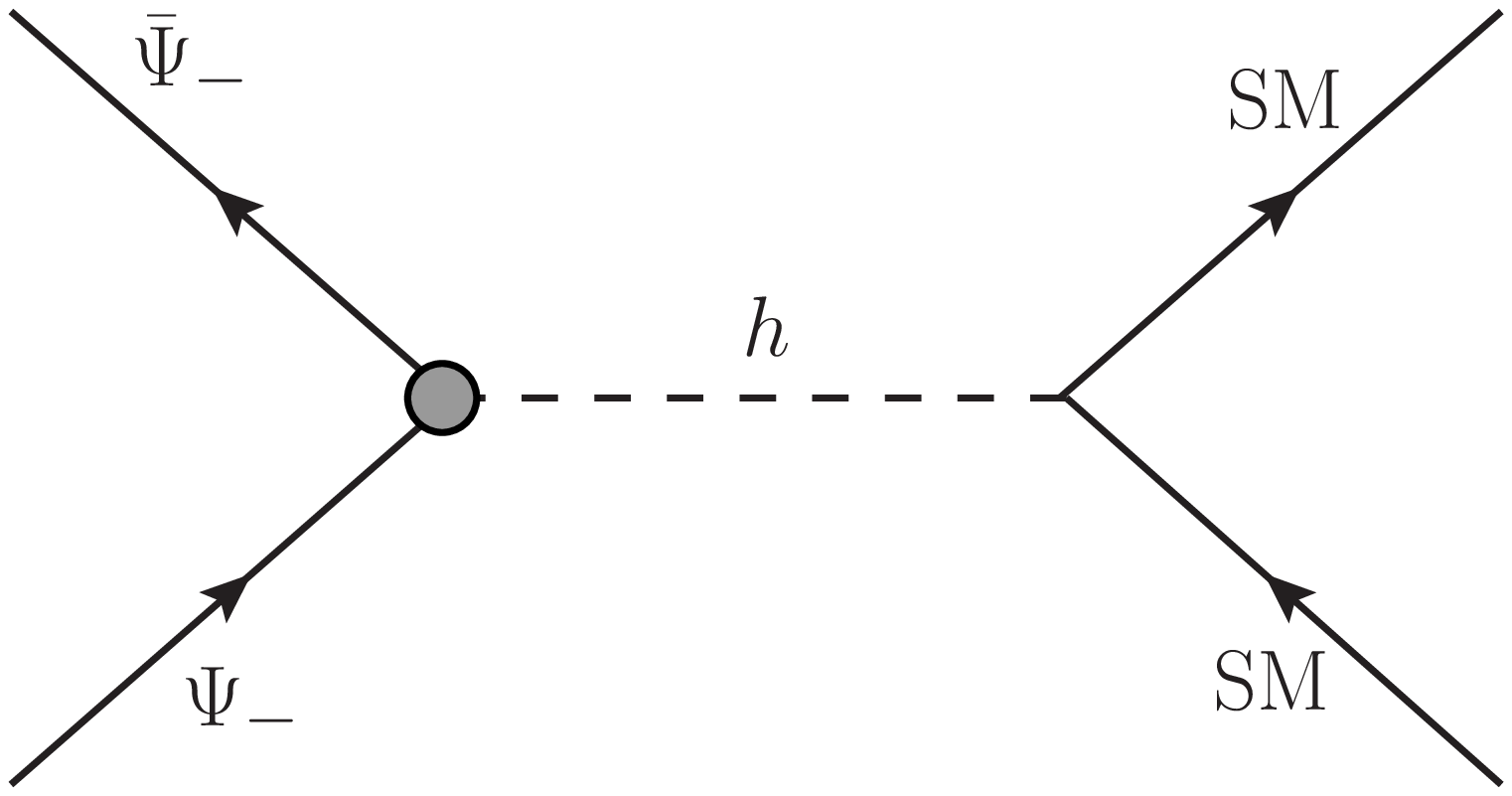} 
  \caption{Pair annihilation of the light state $ \Psi_-$ to the chiral fermion $\chi_5$ and the SM particles. Effective vertex in the last diagram includes mixing of $\phi_i$'s with $h$. And the ``SM'' collectively denotes all massive standard model particles.}
  \label{fig;psiannih}
\end{figure}

It is interesting to realize that the model provides a WIMP-like dark matter particle. Such a particle lies in the TeV heavy fermion sector $\chi_1 - \chi_4$. The fermions $\chi_i$ with $i=1-4$ have no interactions with the SM particles at least to dimension-6 terms in the Lagrangian. Furthermore, they cannot have gauge invariant fermionic bilinear terms $\chi_i \chi_5 f(\phi_j)$ with $f$ being any polynomial of $\phi_j$. Therefore, in terms of mass eigenstates of $\chi'_1 - \chi'_4$, or $\Psi_+$ and $\Psi_-$, the lighter one is stable. The heavy Dirac fermion $\Psi_+$ decays into lighter particles rapidly, such as $\Psi_+ \to \Psi_- \Psi_- \bar{\Psi}_-$ once kinematically allowed, as can be seen from eqs.~\eqref{eq:inteact} and ~\eqref{newspot}. Once upon a time in the early Universe when the temperature was below TeV, after gauge symmetry breaking, the thermal equilibrium contained SM particles and U(1)$'$ sector lightest ones, $\chi_5$ and $\Psi_-$. $\chi_5$ connects the SM sector via its mixing with SM neutrinos, and $\Psi_-$ through the higgs portal, whereas $\chi_5$ and $\Psi_-$ do not have direct interaction. Feynman diagrams for $\Psi_-$ in the thermal equilibrium are drawn in Fig.~\ref{fig;psiannih}.  
As the Universe cooled down to the temperature below $\Psi_-$ mass, the inverse annihilation processes became inefficient and $\Psi_-$ abundance decreased sharply, then due to Hubble expansion the abundance in comoving volume freezes out. The $\Psi_-$ relic abundance today is given by~\cite{KolbAndTurner, DodelsonAndSchmidt}
\begin{equation}
\Omega_{\Psi_-} h^{2}=0.1\left(\frac{x_{f}}{10}\right)\left(\frac{g_{*} }{100}\right)^{1 / 2} \frac{1 \times 10^{-26} \mathrm{~cm}^{3} \mathrm{s}^{-1}}{\langle\sigma v\rangle} \lesssim 0.1,
\end{equation}
where $x_f = m_-^\chi/T_f$ is the mass-to-temperature ratio at the freeze out, which happens at about $x_f \sim 10$, $g_* $ is the effective number of degrees of freedom at $T=  m_-^\chi$, and $\langle\sigma v\rangle$ is the thermally averaged cross section. In this model, the cross section is typical weak interaction alike. The coupling constants are taking natural values $\sim 0.1$ and mediator masses are about TeV. Therefore, $\Psi_-$  plays the same role as WIMP dark matter if the cross section $\langle\sigma v\rangle \gtrsim  10^{-26} \mathrm{~cm}^{3} \mathrm{s}^{-1}$. Note that the last diagram in Fig.~\ref{fig;psiannih} may have a somewhat larger contribution to the cross section, because the mediator mass is smaller, and the number of final states is more than that of other diagrams.  However, the effective vertex includes small scalar mixing, and the cross section also depends on values of new coupling constants.  In this work,we do not distinguish contributions of diagrams in Fig.~\ref{fig;psiannih}.  Nevertheless, there is a natural WIMP-like dark matter in this model.  

As far as the new gauge sector is concerned, the U(1)$'$ gauge boson mass is generated after the spontaneous breaking of the symmetry, it is given by
\begin{equation}
	M_{A'} =  g' \left( \sum_i \tilde{z}^2_i u^2_i \right)^{\frac{1}{2}},
\end{equation}
where $u_i/\sqrt{2}=\left|  \langle \phi_i \rangle \right|$. Due to the kinetic mixing in eq.~\eqref{eq;mixingu1&u1p}, after electroweak and U(1)$'$ symmetries breaking,  the gauge bosons $A_\mu$,  $W^3_\mu$ and $A'_\mu$ connect to the physical gauge bosons  $\hat{A}_\mu$, $Z_\mu$ and $\hat{A}'_\mu$, which are, respectively, the photon, the $Z$ boson an the dark photon. The rotation matrix between these two sets of states are given by 
~\cite{Babu:1997st,deGouvea:2015pea}
\begin{eqnarray}
	\begin{pmatrix}
		\hat{A}_\mu \\
		Z_\mu \\
		\hat{A}^\prime_\mu
	\end{pmatrix} = 
    \begin{pmatrix}
	 \cos \theta_W &  \sin \theta_W & -2 \epsilon \cos \theta_W \\
	 -\sin \theta_W \cos \xi& \cos \theta_W \cos \xi &2\sin \theta_W \epsilon \cos \xi  + \sqrt{1-4\epsilon^2}\, \sin \xi\\
	 \sin \theta_W \sin \xi & -\cos \theta_W \sin \xi &  \sqrt{1-4\epsilon^2}\, \cos \xi -2 \epsilon \sin \theta_W \sin \xi
    \end{pmatrix}
	\begin{pmatrix}
	A_\mu \\
	W^3_\mu \\
	A^\prime_\mu
\end{pmatrix},
\end{eqnarray}
where the angle $\xi$ is defined as $\tan 2\xi =\frac{4\epsilon M^2_Z \sin \theta_W}{M^2_{A'}-m^2_Z} $, up the first order of $\epsilon$. As one can see, when the kinetic mixing is switched off, $\epsilon =0$, as expected, the $\xi$ vanishes and thus gauge bosons in the SM and U(1)$'$ are separated. Furthermore, this mixing  opens a portal linking the SM and U(1)$'$ sectors. In the original basis, $A_\mu$ and $A'_\mu$ couple respectively to the SM and U(1)$'$ fermionic currents, $Q J^\mu A_\mu+Q' J'^\mu A'_\mu$. Because of the kinetic mixing, gauge boson in one sector can couple to a current in the other,
\begin{equation}
	Q \left(\frac{2\epsilon \cos \xi }{\sqrt{1-4\epsilon^2}}+\sin \theta_W\sin \xi \right)J^\mu \hat{A}'_\mu + Q' \frac{\sin \xi}{\sqrt{1-4\epsilon^2}} J'^{\mu} Z_\mu .
\end{equation}
Strength of the above interactions are of the order of $\epsilon$ at the leading order. Depending on how large the $\epsilon$ is, the dark photon could be generated as the SM photon, through processes like electron-positron pair annihilation, Drell-Yan process, Bremsstrahlung etc. And it can decay visibly as well as invisibly, with a main contribution to the visible decay width is proportional to $\epsilon^2$. Taking into account of those production and decay mechanisms, several approaches can be used to search for the dark photon in different experiments, such as collider-based experiment and beam dump experiment. The experimental studies can put limits on the parameter $\epsilon$ and dark photon mass. Detailed analysis of these aspects are left to be carried out in a separate work.

The sterile neutrino mainly decays into three active neutrinos via a $Z$ boson exchange. The decay rate is 
\begin{equation}
	\begin{aligned}
		\Gamma\left(\nu_s \to 3\nu  \right)  = \frac{\sqrt{2}G_F g^2_2\left|U_{i 4}  \right|^2}{12(8\pi)^3 M^2_{W}}m_s^5 ,
	\end{aligned}
\end{equation}
where $g_2$ is SU(2) gauge coupling. The lifetime of $\nu_s$ is about $10^{37}$ s. A more detailed consideration for the lifetime is presented in the appendix~\ref{app:svdecay}. A stable sterile neutrino $\chi_5$ retaining a sizable mixing with active neutrinos inevitably confronts with constraints from cosmological observations. The problem is that the sterile neutrino with an eV-mass and an $\mathcal{O} (0.1)$ mixing with active neutrinos would keep it in the thermal equilibrium in the early Universe. This would increase effective relativistic degrees of freedom, $N_{\mathrm{eff}}$, by one unit, and would modify primordial element abundances during the Big Bang Nucleosynthesis (BBN), and would also affect Cosmic Microwave Background radiation (CMB) anisotropies. The BBN constraint is $N_{\mathrm{eff}} = 2.88 \pm 0.27$ at the 95$\%$ confidence level~\cite{Pitrou:2018cgg}.  The recent CMB result, combining with baryon acoustic oscillation measurements, shows that $N_{\mathrm{eff}} = 2.99 \pm 0.34$ at the 95$\%$ confidence level \cite{Planck:2018vyg}. How to circumvent this problem, per se, is an important task. So far there have been many solutions put forward to alleviate this tension, among which a relevant one to our model is to use an effective 4-point self-interaction of the sterile neutrino, resulting from the integrating out the U(1)$'$ gauge boson \cite{Hannestad:2013ana, Dasgupta:2013zpn}. The solution based on the sterile neutrino self-interaction mechanism can suppress the active-sterile neutrino oscillation and reduces $N_{\mathrm{eff}}$ until to the active neutrino freeze-out temperature, yet at the lower temperature the active-to-sterile neutrino conversion is no longer suppressed and it faces a severe constraint from the CMB bound on the sum of neutrino masses \cite{Mirizzi:2014ama, Chu:2015ipa, Forastieri:2017oma, Chu:2018gxk}. 

We make use of an idea in Refs. \cite{Zhao:2017wmo, Farzan:2019yvo}. By introducing an ultra-light scalar field $\phi$, a very weak Yukawa coupling (of the order of $10^{-23}$) of the sterile neutrino with $\phi$ is assumed. The coupling induces a large sterile neutrino effective mass in the early Universe. This suppresses the mixing between the active and sterile neutrinos. In later times, the scalar field oscillates and the sterile neutrino restores its lightness. In our model, $\phi$ is charged under the U(1)$'$ symmetry. We assign a U(1)$'$ charge $-10/3$ to $\phi$. This allows for gauge invariant interaction $\displaystyle\frac{\zeta}{M^2}\chi_5\chi_5\phi^2_1 \phi$, the very small coupling between $\phi$ and  the sterile neutrino can be achieved for that $ \left|  \langle \phi_1 \rangle \right|\sim 1$ TeV, $M\sim 10^{13}$ GeV and the natural value of the coupling constant $\zeta$. 

 As the last comment, this solution brings in an ultra-light dark matter candidate. The whole model has two main dark matter components. The WIMP-like dark matter abundance can be easily lowered an order of magnitude by adjusting the model parameters, in order to make the total dark matter abundance consistent with observations, without changing the qualitative property of the model. 

\section{C\lowercase{oncluding remarks}} 
\label{sec:Concl}

In summary, for the sterile neutrino, we have extended the SM by including in a U(1)$'$ chiral sector, and the new sector is basically fixed by anomaly free conditions. The eV-scale mass for the sterile neutrino and its mixing with three active neutrinos are naturally generated. This model includes some interesting phenomenology. Let us highlight the model with following remarks:

(i) In our model, the sterile neutrino, as a chiral fermion, has a default vanishing mass under the local gauge symmetry.  Even after the symmetry breaking, our choice of the particle content still keeps the $\chi_5$ massless until to dimension-5 operators.  The small mass is therefore explained. In addition, the other 4 new chiral fermions happen to make a pair of Dirac particles.   

(ii) The decay lifetime of the sterile neutrino has been computed. It is a very stable particle. The model automatically gives a WIMP-like Dirac fermion dark matter candidate that (together with an ultra-light scalar field necessary for explanation of the cosmological constraints) can fulfill the dark matter relic abundance observed today.

(iii) We have taken all the new coupling constants to be in natural ranges, while unable to explain unnatural values of those in the SM.  This is a simplified way in the analysis.  The new physics scale $M$ is the one at which the SM sector and the U(1)$'$ sector unify.  It is interesting to notice that $M$ is also the seesaw scale \cite{Minkowski:1977sc, Yanagida1979, Gell-MannRamon, Glashow1980dfc, Mohapatra:1979ia}. While in low energies the SM sector and the U(1)$'$ sector are separated except for a connection through the higgs portal.

(iv) Our mechanism of generating a light chiral fermion beyond the SM is of generic features. We expect that it is not restricted to the eV-scale sterile neutrino, it can be applied in other physics scenarios about new fermions, such as the warm dark matter case.

%

\section*{A\lowercase{cknowledgements}}
The authors acknowledges support from the National Natural Science Foundation of China (No. 11875306) and the Key Research Program of the Chinese Academy of Sciences (No. XDPB15).


\appendix

\section{the light sterile neutrino decay}
\label{app:svdecay}

\begin{figure}[ht]
	\includegraphics[width=0.45\textwidth]{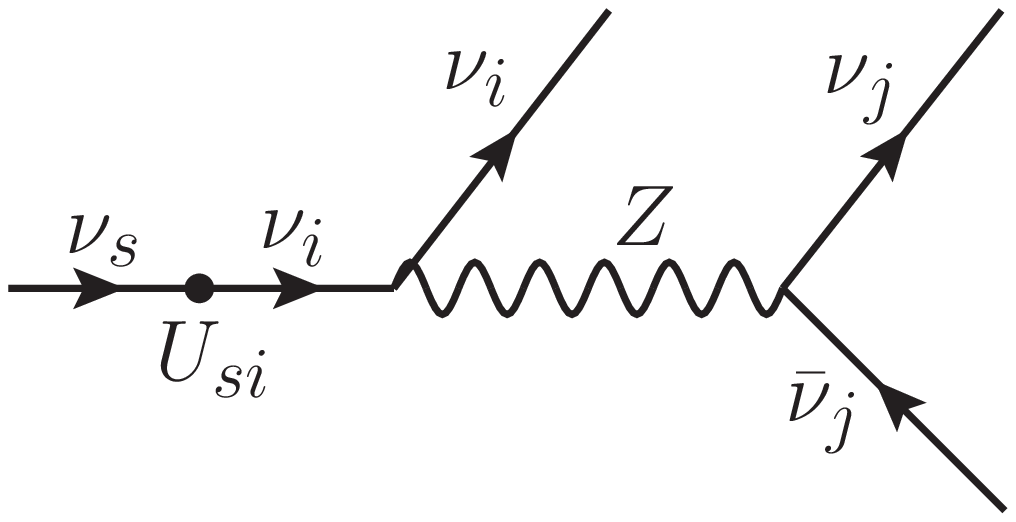} 
	\includegraphics[width=0.45\textwidth]{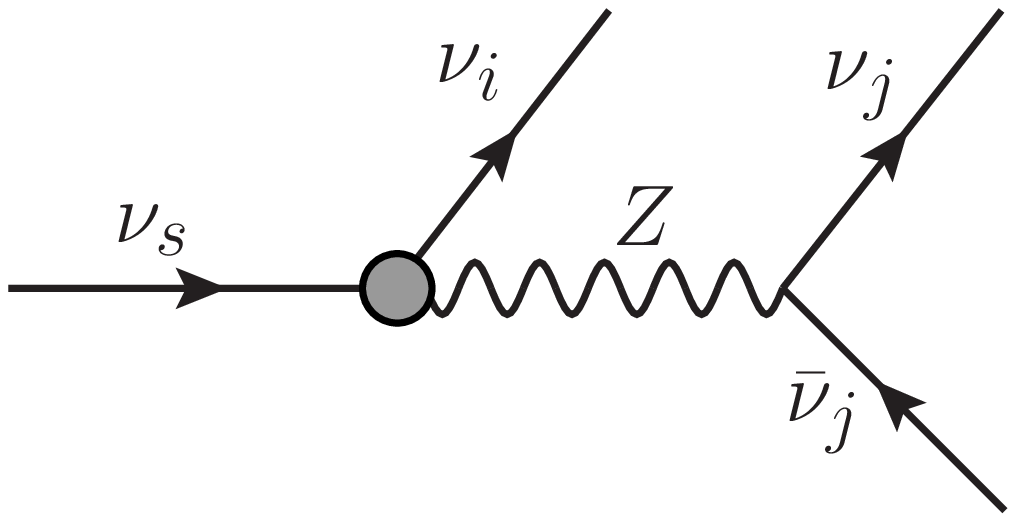}
	\caption{
		The Feynman diagrams for the sterile neutrino decay into three active neutrinos. Effective vertex illustrated with a blob takes into account of both neutrino mixing and gauge boson mixing.}
	\label{fig;3nusdecay}
\end{figure}

The sterile neutrino has two decay channels. It can decay into three active neutrinos via $Z$ boson mediation. The Feynman diagrams are shown in Fig.~\ref{fig;3nusdecay}, where a dot in the first diagram is for active-sterile neutrino mixing and a blob vertex in the second diagram includes both active-sterile neutrino mixing and gauge field mixing. 
We compute the decay rate of this process, 
\begin{equation}
	\begin{aligned}
		\Gamma\left(\nu_s \to 3\nu  \right)  = \frac{\sqrt{2}G_F \left[g_2 + g' \sin(2\theta_{\rm W})   \epsilon \right]^2\left|U_{i 4}  \right|^2}{12(8\pi)^3 M^2_{W}}m_s^5 ,
	\end{aligned}
\end{equation}
where $g_2$ is SU(2)$_L$ gauge coupling, $\theta_\text{W}$ is weak mixing angle.
Exploiting values of known parameters, the above rate yields
\begin{equation}
	\begin{aligned}
		\Gamma\left(\nu_s \to 3\nu  \right)  = & \ 1.34\times 10^{-61} \left[\,g+ g'\sin(2\theta_{\rm W}) \epsilon\, \right]^2 \left(\frac{m_s}{1\mathrm{eV}} \right)^5 \mathrm{GeV},\\
		\tau_{\nu_s} = & \ 4.91 \times 10^{36}  \left[\,g+ g'\sin(2\theta_{\rm W}) \epsilon\, \right]^{-2} \left( \frac{1\; \mathrm{eV}}{m_s}\right)^5 \mathrm{s}\, .
	\end{aligned}
\end{equation}
Taking $g_2=0.65$, $g'\simeq 0.35$ as that of the SM $U(1)_\mathrm{Y}$ gauge coupling, $\sin^2 \theta_{\rm W}=0.23$, and $\epsilon \simeq 10^{-3}$, value of the decay rate is $\Gamma\left(\nu_s \to 3\nu  \right)\approx  5.67 \times 10^{-62} \; \mathrm{GeV}$, the lifetime of the sterile neutrino is $\tau_s = 1.16\times 10^{37} \; \mathrm{s}$.

\begin{figure}[ht]
	\includegraphics[width=0.45\textwidth]{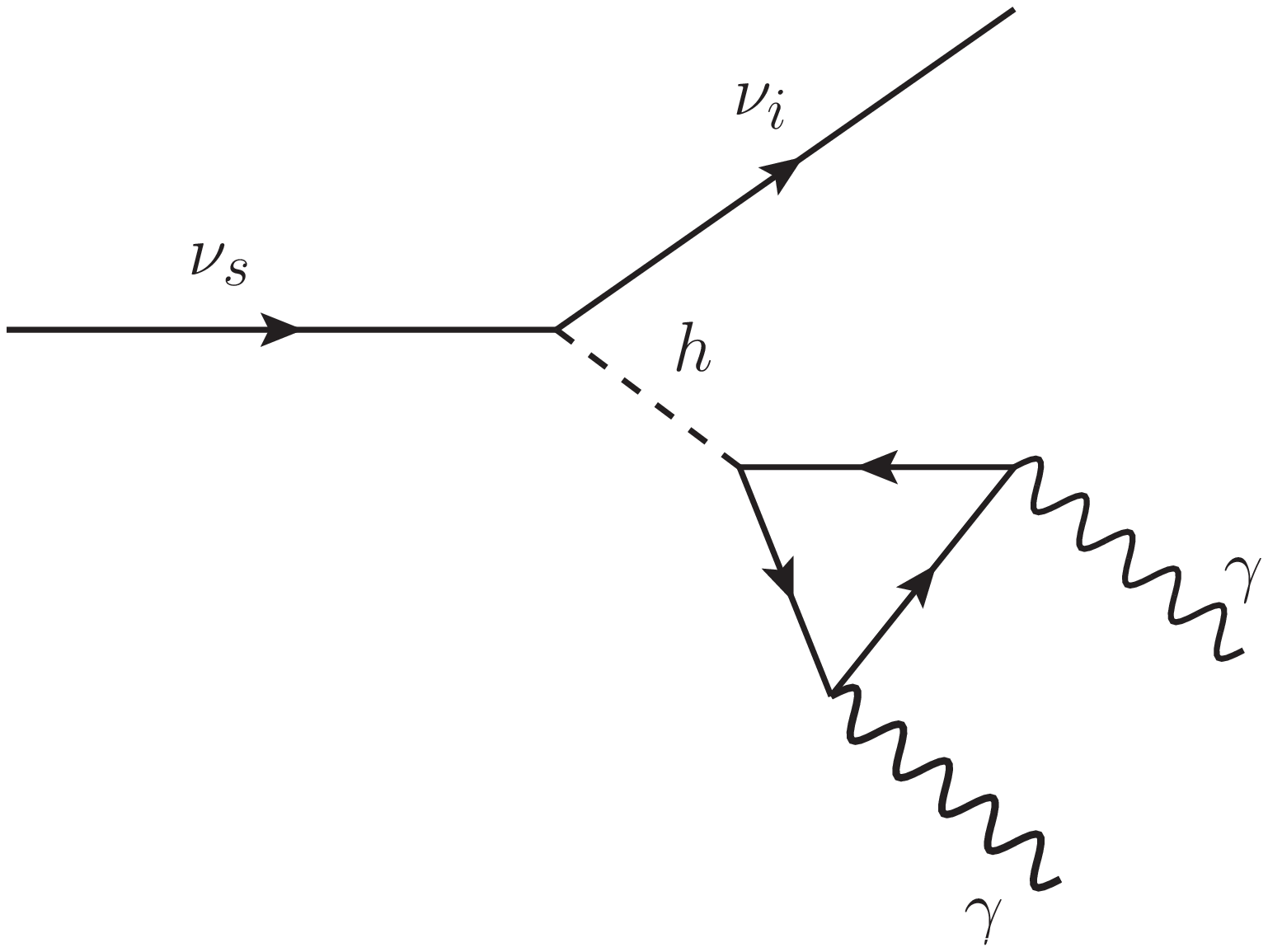} \hspace*{2mm}
	\includegraphics[width=0.45\textwidth]{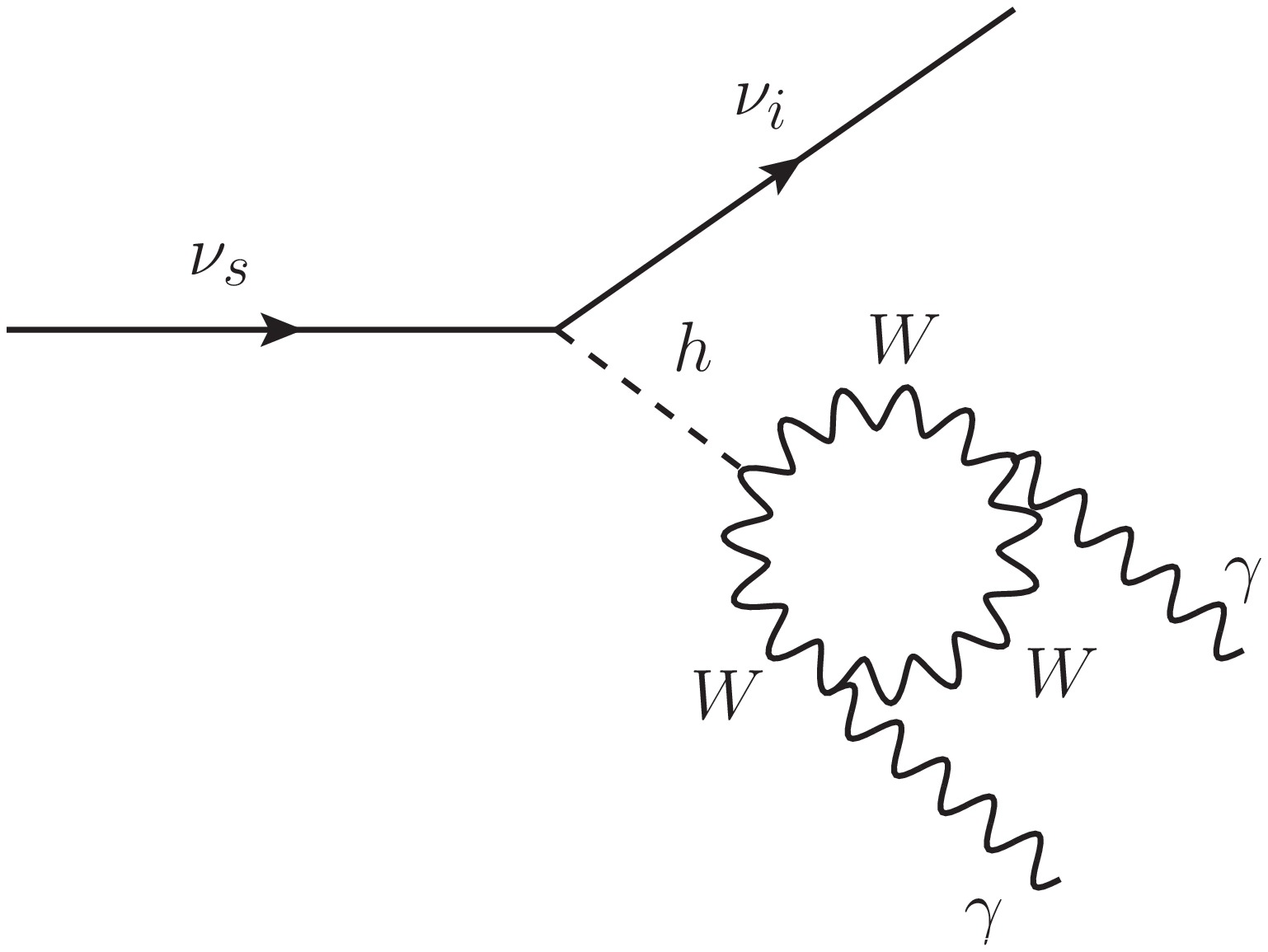}\\
	\vspace*{5mm}
	\hspace*{-0.47\textwidth}
	\includegraphics[width=0.45\textwidth]{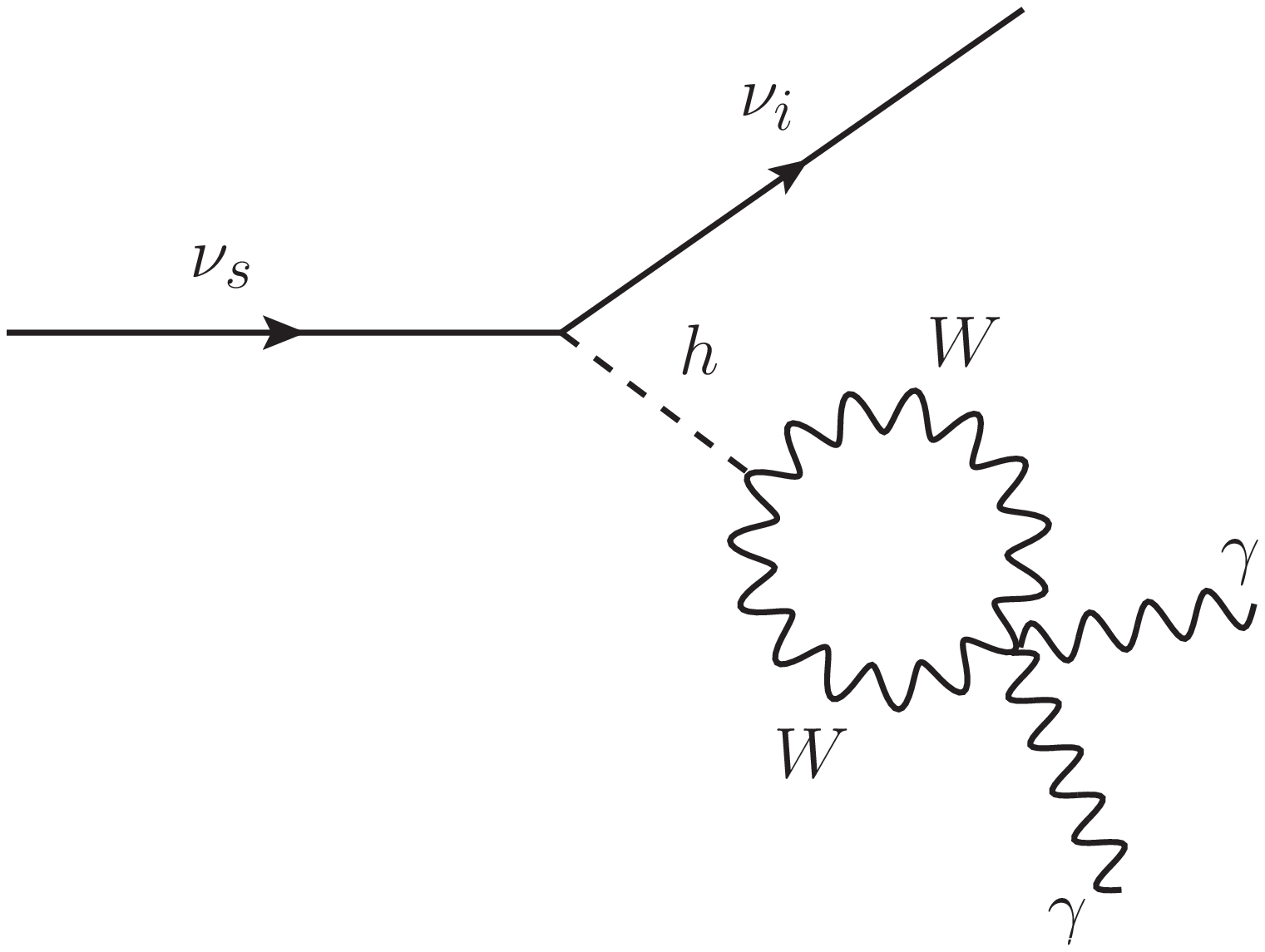}
	\caption{
		Feynman diagrams for the sterile neutrino decay $\nu_s \to \nu_i \gamma \gamma$ in the unitary gauge. Here we did not show another two diagrams, which result from the exchange of two photon vertices, since they contribute the same scattering amplitudes as above.}
	\label{fig;nusdecay}
\end{figure}

There exists another subdominant decay channel, which may be of academic interest, at least it is specific to this model. The sterile neutrino can decay into an active neutrino and two photons. The coupling in  eq.~\eqref{eq;mixingu1&u1p} opens up this possibility, as shown in Fig.~\ref{fig;nusdecay}.  
The Lagrangian describing the $h \gamma\gamma$ effective vertex can be written as \cite{Ellis:1975ap, Zheng:1990qa, Marciano:2011gm}
\begin{equation}
	\mathcal{L}^{h\gamma\gamma}_{\mathrm{eff}}=-\frac{\alpha L(\lambda_p ) }{2 \pi v}  F_{\mu \nu} F^{\mu \nu} h, 
\end{equation}
where $\alpha$ is the fine structure constant, $v = 246 \; \mathrm{GeV}$, $\lambda_p$ is the mass squared ratio of particle $p$ running in loops, $\lambda_{p}=m_{p}^{2} / m_{h}^{2}$, and $L(\lambda_p )$ is a function obtained from loop integrals. Since dominant contributions come from $W$ boson and top quark loops, $L(\lambda_p ) \simeq 3/2$. We obtain the sterile neutrino decay rate,
\begin{equation}
	\begin{aligned}
		\Gamma\left(\nu_s \to \nu \gamma \gamma  \right)  = & \ \frac{2m^7_s}{15(8\pi)^3}\left(  \frac{\alpha}{\pi v m^2_h} \right)^2  \left| L(\lambda_p )\right|^2 \, \sum_{i} \left| C_i\right|^2 \\
		= & \ 6.9\times 10^{-87} \left( \sum_{i} \left| C_i\right|^2 \right) \left(\frac{m_s}{1\mathrm{eV}} \right)^7 \mathrm{GeV},,
	\end{aligned}
\end{equation}
by neglecting the active neutrino masses. Here $C_i=\lambda^i_3 \langle \phi_3\rangle/M$ which measures the coupling strength among the sterile neutrino, the active neutrino and the higgs boson.

It is clear that the decay rate of $\nu_s \to 3\nu$ is much larger than that of $\nu_s \to \nu \gamma \gamma$ channel. This is due to the fact that there is no loop suppression in the former one. As an estimate for the result of the loop process, let us take $0.1$, $1$ TeV, $10^{14}$ GeV for values of $\lambda^i_3$, $\langle \phi_3\rangle$, $M$ respectively, this gives $\left| C_i\right| = 10^{-12} $, thus $\Gamma\left(\nu_s \to \nu \gamma \gamma  \right) \approx 6.21 \times 10^{-110}\;  \mathrm{GeV}$, the corresponding lifetime is $\tau_{\nu_s} \approx 1.06 \times 10^{85}$ s. This is negligible compared to the results from  $\nu_s \to 3\nu$ channel.
In either decays, the lifetime of the sterile neutrino is much longer than age of the Universe. Therefore, the sterile neutrino is an extremely stable particle.


\providecommand{\href}[2]{#2}\begingroup\raggedright\endgroup

\end{document}